\documentclass[11pt, letterpaper]{article} 

\usepackage[includeheadfoot,
            marginratio={1:1,2:3}, 
            width=440pt, 
            height=688pt,]{geometry}

\usepackage{amsmath}
\usepackage{amsfonts}
\usepackage{amssymb}
\usepackage{bbm,bm}
\usepackage{mathtools}
\usepackage{slashed}
\usepackage{mathalfa}
\usepackage{bbold}

\usepackage{graphicx,caption,subfigure}
\usepackage{tikz} 
\usepackage{tikz-cd}
\usetikzlibrary{decorations.pathreplacing,calc,tikzmark}
\usepackage{booktabs}
\usepackage{adjustbox}
\usepackage[utf8]{inputenc}
\usepackage[splitrule]{footmisc} 
 
\usepackage{placeins}
\usepackage{empheq}
\usepackage{paralist}
\usepackage{cite}
\usepackage[normalem]{ulem}
\usepackage{soul}

\usepackage[colorlinks=false,urlbordercolor=red
]{hyperref}
\usepackage{xcolor}
\usepackage{tensor}

\newcommand{\nc}{\newcommand}
\nc{\lb}{\llbracket}
\nc{\rb}{\rrbracket}
\nc{\gl}{\llbracket}
\nc{\gr}{\rrbracket}
\nc{\del}{\partial}
\nc{\eq}[1]{\begin{equation}
                     \begin{split} #1 \end{split}
                     \end{equation}
}
\nc{\ov}{\overline}
\nc{\uv}{\underline}
\nc{\fa}{\hat}
\nc{\fb}{\MakeUppercase}
\nc{\fc}{\tilde}
\nc{\myhash}{\raisebox{\depth}{\#}}

\allowdisplaybreaks[2]
\numberwithin{equation}{section}
\DeclareUnicodeCharacter{2212}{-}

\begin{document}


\vspace*{-1.5cm}
\begin{flushright}
  {\small
  MPP-2025-51\\
   LMU-ASC 08/25
  }
\end{flushright}

\vspace{1.0cm}
\begin{center}
  {\huge Non-associative Algebras of Cubic Matrices \\[0.3cm] 
    and their Gauge Theories} 
\vspace{0.4cm}

\end{center}

\vspace{0.25cm}
\begin{center}
{
  \Large Ralph Blumenhagen$^{1}$, Antonia Paraskevopoulou$^{1,2}$ and\\[0.2cm]
   Thomas Raml$^{1}$

}
\end{center}

\vspace{0.0cm}
\begin{center} 
  \emph{$^{1}$ 
Max-Planck-Institut f\"ur Physik (Werner-Heisenberg-Institut), \\ 
Boltzmannstra\ss e  8,  85748 Garching, Germany } 
\\[0.1cm] 
\vspace{0.25cm} 
\emph{$^{2}$ Fakult{\"a}t f{\"u}r Physik, Ludwig-Maximilians-Universit{\"a}t M\"unchen, \\ 
Theresienstr.~37, 80333 M\"unchen, Germany}\\[0.1cm]
\vspace{0.3cm}
\end{center} 
\vspace{0.5cm}

\begin{abstract}
  Motivated by M-theory, we define a new type of non-associative
  algebra involving  usual and cubic matrices at the same time. The resulting algebra can be regarded as a  two-term truncated $L_\infty$ algebra giving rise to a fundamental identity between the two- and the three-bracket. We provide a simple class of concrete examples of such algebras based on the structure constants of a Lie algebra. Connecting to previous results on higher structures, we generalize the construction of Yang-Mills theories, topological BF theory and generalized IKKT models
and point out some appearing issues.
\end{abstract}

\thispagestyle{empty}
\clearpage

\setcounter{tocdepth}{2}



\section{Introduction}

The most mysterious aspect  of our current understanding
of the space of consistent quantum gravity theories
is certainly the strong coupling limit
of ten-dimensional type IIA string theory, namely M-theory.
Not only is no complete fundamental theory of M-theory known,
but even its fundamental degrees of freedom 
are obscure.

The most promising formulation we have so far is
certainly the BFSS Matrix Model \cite{Banks:1996vh} (see
\cite{Bilal:1997fy,Bigatti:1997jy,Taylor:2001vb}
for reviews), where the latter is given by the theory of ${\cal N}$ $D0$-branes and their
matrix model interactions.
By studying the interactions of bound states of such $D0$-branes \cite{Kabat:1997im,Kabat:1997sa},
one could detect that this theory also contains a longitudinal
and a transverse $M2$-brane current.
However, the $M5$-brane current did not occur in the same clear manner
already at finite ${\cal N}$, the reason being that the transverse $M5$-brane current was
just absent and the longitudinal one was vanishing  due to the Jacobi identity for matrices.
The necessity of including the $M5$-branes at a fundamental  level is
also suggested by recent advances in the swampland program,
namely by the M-theoretic emergence proposal (see \cite{Blumenhagen:2024lmo} for a review). 
It turned out that in the decompactification limit to M-theory,
where one keeps the ten-dimensional Planck scale constant,
the species scale is
the eleventh dimensional Planck scale and the lightest towers of states are D0-branes (KK modes),
though with transverse $M2$- and $M5$- branes still at the
species scale \cite{Blumenhagen:2023xmk}. Therefore, in
this limit $M2$- and $M5$-branes appear on equal footing.

Hence, in order to include  $M5$-branes at a fundamental level
the first idea one can have is
to violate the Jacobi identity meaning
that one has to go beyond usual matrices and consider
other objects, which can be non-associative.
This is precisely the direction in which we want to go allowing us to 
connect to previous work on three-brackets and higher structures
(see e.g. the proceedings of \cite{Jurco:2019woz} and citations therein).
That non-trivial three-brackets might be relevant
is certainly not a new idea, as they made a  prominent
appearance already  in the Bagger-Lambert-Gustafson (BLG) theory
\cite{Bagger:2006sk,Gustavsson:2007vu,Bagger:2007jr} of multiple
$M2$-branes (see \cite{Bagger:2012jb} for a review).
Moreover, non-associative structures were also proposed
to arise in non-geometric string backgrounds with $R$-flux \cite{Blumenhagen:2010hj,Lust:2010iy}
(see \cite{Plauschinn:2018wbo} for a review).

The fact that a stack
of $\mathcal{N}$ $M5$-branes has $\mathcal{N}^3$ degrees of freedom \cite{Klebanov:1996un} led to the proposal
\cite{Awata:1999dz,Kawamura:2002yz,Kawamura:2003cw,Ho:2008bn}
that one might introduce cubic matrices as  new degrees
of freedom\footnote{Quartic and higher
  index matrices were considered in \cite{Ashwinkumar:2021hax}.}. 
 In this paper we take a new approach to the implementation of cubic
 matrices in  a physical context.
 Unlike the previous attempts to directly define a ternary product for three such cubic matrices,
 in section 2 we construct an algebra that involves both
 usual ${\cal N}\times {\cal N}$ matrices (bimatrices)
 and cubic ${\cal M}\times {\cal N}\times {\cal N}$ matrices (with ${\cal M}$ not necessarily equal
 to ${\cal N}$) at the same time. Such a construction could be  motivated
 by the  aim of extending  the BFSS matrix model without destroying
 its previous success, which involved, of course, bimatrices.
 
To be more precise, this allows us to naturally define a $\mathbb Z_2$ graded binary
 product between these elements. For instance, the product of
 two cubic matrices will be a bimatrix and the product of a bimatrix
 and a cubic matrix will give  a cubic matrix.
 The product is a natural generalization of the product between
 two bimatrices, but lacks its associativity.
 One way to proceed is to use the fact that such a non-associativity
 gives the algebra the structure of a two-term $L_\infty$
 algebra \cite{Hohm:2017cey},
 shortly denoted as  $L^{\rm cub}_2$ in the sequel.
 Due to its non-associativity, the Jacobi identity is generally not satisfied
 and gives rise to the definition of a three-bracket $[\cdot,\cdot,\cdot]$.
 The combination of  the commutator two-bracket and this three-bracket
 are subject to a fundamental identity, which will
 play the same essential role as the Jacobi identity plays for Lie algebras.
Let us emphasize that the whole structure is not just abstract
 but all operations are in the end defined in terms
 of the above mentioned binary  bi-/cubic matrix multiplications.
 We provide a class of concrete $L^{\rm cub}_2$ algebras,
 which combines arbitrary rank  ${\cal N}\times {\cal N}$ bimatrices with
 ${\cal M}\times {\cal N}\times {\cal N}$ cubic matrices.
 In this respect, we also comment on the difference to
  Lie 3-algebras  which appeared for BGL theories.

 As a first approach, in section 3 we address the question of whether this algebraic structure
 can be consistently implemented in gauge theories,
 a question which has been addressed more abstractly for general 2-term
 $L_\infty$ algebras already in \cite{Zucchini:2011aa,Ritter:2013wpa}.
 In fact, the $L^{\rm cub}_2$ algebras just provide  a concrete example of
 this approach.
  In contrast to usual Yang-Mills theories,
 based on Lie algebras, the non-vanishing of the Jacobi identity
 gives rise to new terms involving the non-trivial three-bracket.
 The whole structure is unavoidably more complicated
 but, as we will show, is still computable with its internal
 consistency governed by the above mentioned fundamental
 identity.
 As a new feature, gauge consistency requires the  introduction of higher form gauge fields. We discuss several possible modifications in order to bypass some of the shortcomings of the resulting theories. Furthermore we briefly review how the $L^{\rm cub}_2$ algebra can provide realizations of topological BF theory and deformed IKKT models. We conclude with some preliminary results on extending the Yang-Mills structure to fermionic matter fields.

 \section{The \texorpdfstring{$L_\infty$}{L∞} algebra of cubic matrices}

We recall that for  a  quantized version of the Nambu-bracket alone
it was suggested \cite{Awata:1999dz,Kawamura:2002yz,Kawamura:2003cw}
that it  can be represented by cubic matrices
$a_{ijk}\in\mathbb C$,
carrying three indices whose entries are complex numbers.
Such an idea is also well motivated by the known fact that
a stack of ${\cal N}$ $M5$-branes supports of the order
${\cal N}^{3}$ degrees of freedom \cite{Klebanov:1996un}.
One defines directly a ternary product $(abc)$ of three cubic matrices. 
This is interesting but allows to evaluate only odd products of cubic matrices.
To really define a complete set of calculation rules for  such cubic matrices, one also needs to define
a binary product for them.
One could try to construct  such a product such  
that the result is  again a cubic matrix.
However, contemplating about a non-associative extension of
the quite successful BFSS matrix model
something else is suggested, namely
to try to formulate a consistent
set of calculation rules for  bimatrices {\it and} cubic matrices.

\subsection{Cubic matrices and their products}

For that purpose, let us consider a vector space $V=V_B\oplus V_C$ of bimatrices $M_{ij}$ and
cubic matrices $a_{rij}$ where $i,j=1,\ldots,{\cal N}$ and, as we will
see, we can allow the index $r$ to run over a different regime, i.e.
$r=1,\ldots,{\cal M}$. In the following, cubic matrices are denoted
by small letters $a,b,c,\ldots$, bimatrices by $M,N,P,\ldots$
and if we are indifferent about the type of object we denote
elements of $V$ as $X_1,X_2,\dots$.

Here we consider hermitian objects, where as usual a bimatrix
is  hermitian if $M_{ji}=M_{ij}^*$ and we call a cubic
matrix hermitian if $a_{rji}=a_{rij}^*$, for all $r=1,\ldots,{\cal
  M}$. This means that the hermitian conjugate of a cubic matrix is
defined via
\eq{
            ( a^\dagger )_{rij}=a^*_{rji}\,.
}
There exist  natural definitions  for the mutual product of two
such objects.
The product of two bimatrices is just the usual matrix product
\eq{
  \label{prodbibi}
        (M\cdot N)_{ij}= \sum_{m=1}^{\cal N}  M_{im} \,N_{mj}\,.
      }
In the same spirit, the product of two cubic matrices
 can be defined to yield a bimatrix via
\eq{
  \label{prodcubiccubic}
          (a\cdot b)_{ij}= \sum_{r=1}^{\cal M} \sum_{m=1}^{\cal N}  a_{rim} \,b_{rmj}\,.
}  
It remains to define the product of a matrix and a cubic matrix for which
there exists a similar natural choice, namely
\eq{
  \label{prodbicubicf}
          ( M\cdot a)_{rij}= \sum_{m=1}^{\cal N}  M_{im}
          \,a_{rmj}\,,\qquad
           ( a\cdot M)_{rij}= \sum_{m=1}^{\cal N}  a_{rim}\, M_{mj}\,.
}  
The question is what kind of structure, if any, these three products define.

It is convenient to represent a cubic matrix as an ${\cal M}$-tupel  of
${\cal N}\times {\cal N}$ bimatrices
\eq{
         a=( a_1,\ldots, a_{\cal M} )\,.
       }
These bimatrices from a ring ${\cal R}$ under matrix addition  and multiplication.
Then the multiplication \eqref{prodbicubicf} is nothing else
than
\eq{
  M\cdot  a=( M a_1,\ldots, M a_{\cal M} )\,,\qquad
   a\cdot  M=( a_1 M,\ldots, a_{\cal M} M )
       }
 for $M\in {\cal R}$. This multiplication identifies  the space of cubic matrices as
 a left-right module over the ring ${\cal R}$.
 Now one could define the tensor product $a\otimes b$ of two such ${\cal M}$ tuples
which is an object with four indices given by
\eq{
  \label{tensorfourindex}
      (a\otimes b)_{rs}=a_r\, b_s \in {\cal R}\,,\qquad\quad  (a\otimes b)_{rs,ij}= \sum_{m=1}^{\cal N} a_{rim} \,b_{smj}\,.
}
Via iteration one could generate higher index objects.
This could be an interesting direction to follow but here
we just observe that  the product \eqref{prodcubiccubic} between two cubic matrices
can be regarded  as the  trace of the tensor product, namely
 \eq{
     a\cdot b= \sum_{r=1}^{\cal M} (a\otimes b)_{rr} =\sum_{r=1}^{\cal M} a_r\, b_r \in {\cal R}\,.
  } 
 As we will see in section \ref{sec_SU2ex}, this identification  of cubic matrices
 as elements in a module is helpful in
 doing concrete computations.

We also observe that  the products respect a $\mathbb Z_2$ grading
\eq{
           M\cdot M \to M\,,\qquad  a\cdot  a\to M\,,\qquad  M\cdot  a \to a\,,\qquad  a\cdot  M \to a\,,
}
where the bimatrices  $M\in V_B$ carry even degree and the cubic
matrices $a\in V_C$ odd degree.
While the usual product between two bimatrices \eqref{prodbibi}
is associative, generically the associator for three cubic matrices
is non-vanishing, i.e.
\eq{
                              (a\cdot b)\cdot c\ne a\cdot (b\cdot c)\,.
}         
However, one can straightforwardly
check that all associators involving at least one bimatrix still vanish, i.e
\begin{equation}\label{associatornull}
\begin{aligned}
  (Ma)b-M(ab)&=0\,, &\qquad (aM)b-a(Mb)&=0\,, &\qquad  (ab)M-a(bM)&=0\,,\\
  (MN)a-M(Na)&=0\,, &\qquad (Ma)N-M(aN)&=0\,, &\qquad (aM)N-a(MN)&=0\,.
\end{aligned}
\end{equation}
Hence, defining the  commutator between two elements of $V$ as
\eq{
  \label{commute}
  [X_1,X_2]=X_1\cdot X_2-X_2\cdot X_1\,,
}  
all Jacobiators
\eq{
{\rm Jac}(X_1,X_2,X_3):=[[X_1,X_2],X_3]+[[X_2,X_3],X_1]+[[X_3,X_1],X_2]
}
involving at least one bimatrix do vanish.
Note that we always take the commutator, i.e.
in contrast to super Lie algebras the definition
of the bracket does not involve the $\mathbb Z_2$ grades of $X_1$ and $X_2$.

The trace of  a bimatrix is defined as usual and it turns out to be reasonable
to define the trace of any cubic matrix to vanish, i.e.
\eq{
  {\rm tr}(M)=\sum_{i=1}^{\cal N} M_{ii}\,, \quad\qquad {\rm tr}(a)=0\,.
  }
This implies that only  even $\mathbb Z_2$ objects can have a
non-vanishing trace.
Then for both bi- and cubic matrices one can define a positive
definite inner product as
\eq{\label{Z2innprod}
  \langle M_1,M_2\rangle= {\rm tr}(M_1^\dagger\, M_2)\,,\qquad
   \langle a_1,a_2\rangle= {\rm tr}(a_1^\dagger\, a_2)
}  
so that
\eq{
  \langle M,M\rangle=\sum_{i,j} |M_{ij}|^2> 0\,,\qquad
  \langle a,a\rangle=\sum_{r,i,j} |a_{rij}|^2> 0\,
}
for $M\ne 0$ and $a\ne 0$.
Moreover, besides the cyclicity property of the trace for products of  bimatrices,
one has 
\eq{
  \label{traceprop}
              {\rm tr}(M a b)=   {\rm tr}(a b M) =  {\rm tr}(b M a) \,,
}
where the  first equality follows from the cyclicity for  bimatrices
and the second one can readily be confirmed from the definitions
\eqref{prodbibi},\eqref{prodcubiccubic} and \eqref{prodbicubicf}
of the products.

\subsection{Cubic matrices and a 2-term \texorpdfstring{$L_\infty$}{L∞} algebra}
\label{sec_cubic}

It was generally shown in \cite{Hohm:2017cey} that such a structure,
i.e. an antisymmetric bracket not necessarily satisfying the Jacobi
identity, can always be extended to a 2-term $L_\infty$ algebra.
In this section we verify this for our case.

In general, an $L_\infty$ algebra consists of a graded vector space
$V=\oplus_n V_n$ equipped with multi-linear products
$\ell_n(X_1,\ldots,X_n)$ satisfying quadratic relations.
We have delegated more details about 
 the formal definition of
an $L_\infty$ algebra to appendix \ref{app_linfty}.
For our concrete case, the first step is to define a two-term graded vector space\footnote{Note that
  the grading in the definition of the $L_\infty$ algebra is
  not the $\mathbb Z_2$ grading we introduced in the vector space $V$
  of bi- and cubic matrices.}
\eq{
  \label{linftytwoterm}
                           V=V_0\oplus V_1 
}
with all other $V_k$ vanishing.
Next  we choose $V_0=V_1=V_B\oplus V_C$, i.e. both contain bimatrices $M_i$ and cubic
matrices $a_m$. Let us notationally distinguish elements from $V_0$ and $V_1$
by denoting them as $X_i$ and $Y_i$, respectively.
Moreover, $\ell_1$ acts like the identity on $V_1$
and gives zero when acting on any element from $V_0$, i.e.
\eq{
  \ell_1(Y_i)=X_i\,,\qquad  \ell_1(X_i)=0\,.
}
This guarantees  that the first $L_\infty$ relation from
\eqref{linftyrela}, namely
${\cal J}_1=\ell_1\ell_1=0,$ is trivially satisfied.
Next we define $\ell_2$ as
\begin{equation}
\begin{aligned}
  \ell_2(X_1,X_2)&:= [X_1,X_2]\in V_0\,, &\qquad  \ell_2(X_1,Y_2)&:= [X_1,Y_2]\in V_1\,,\\
  \ell_2(Y_1,Y_2)& :=0\,, &\qquad   \ell_2(Y_1,X_2)&:= [Y_1,X_2]\in V_1\,,
\end{aligned}
\end{equation}
where the bracket $[\cdot,\cdot]$ means the commutators \eqref{commute}
between bi- and cubic matrices.
Then, the relations ${\cal J}_2=0$ are only non-trivial for ${\cal J}_2(X_1,Y_2)\in V_1$ and read
\eq{
            \ell_2(\ell_1(X_1),Y_2)+\ell_2(X_1,\ell_1(Y_2))=\ell_1(\ell_2(X_1,Y_2))\,.
}
Using that $\ell_1(X_1)=0$ both sides are equal to $[X_1,Y_2]$ so that
the relation is indeed satisfied.

The relation ${\cal J}_3=0$ \eqref{ininftyrel2} is only non-trivial for ${\cal J}_3(X_1,X_2,X_3)\in V_0$
and ${\cal J}_3(X_1,X_2,Y_3)\in V_1$ (and its permutations). The first can be satisfied
by introducing a non-trivial three-bracket $\ell_3$ via
\eq{
  \label{defthreebracket}
             \ell_3(X_1,X_2,X_3):=-\Big( [[X_1,X_2],X_3]+[[X_2,X_3],X_1]+[[X_3,X_1],X_2]\Big)\in V_1\,.
}
Due to the associativity relations \eqref{associatornull} this is only non-vanishing  for three cubic matrices
so that    $\ell_3(a_1,a_2,a_3)\ne 0$. Automatically, this $\ell_3$ also trivializes
the second condition ${\cal J}_2(X_1,X_2,Y_3)=0$.

Since we have a non-trivial $\ell_3$, also the next condition  ${\cal J}_4=0$  matters.
Here the only in principle non-trivial combination is  ${\cal J}_4(X_1,X_2,X_3,X_4)\in V_1$,
which in total reads
\eq{
  \label{jfourtotal}
  &\ell_2\big(\ell_3(X_1,X_2,X_3),X_4\big)-\ell_2\big(\ell_3(X_2,X_3,X_4),X_1\big)+\ell_2\big(\ell_3(X_3,X_4,X_1),X_2\big)\\
  & -\ell_2\big(\ell_3(X_4,X_1,X_2),X_3\big)=
  \ell_3\big(\ell_2(X_1,X_2),X_3,X_4\big)-\ell_3\big(\ell_2(X_2,X_3),X_4,X_1\big)\\
  &+\ell_3\big(\ell_2(X_3,X_4),X_1,X_2\big)-\ell_3\big(\ell_2(X_4,X_1),X_2,X_3\big)-\ell_3\big(\ell_2(X_1,X_3),X_2,X_4\big)\\
  &-\ell_3\big(\ell_2(X_2,X_4),X_1,X_3\big)\,.
}
Defining the three-bracket via the Jacobiator \eqref{defthreebracket},
this relation is automatically satisfied (like the usual
Jacobi identity for a commutator of associative objects).
In our special case, this relation reduces considerably.
Due to the associativity relations \eqref{associatornull}, the only non-trivial
combinations are   ${\cal J}_4(a_1,a_2,a_3,M)$ and ${\cal J}_4(a_1,a_2,a_3,a_4)$.
In the first case, the condition \eqref{jfourtotal} reduces to
\eq{
  \label{newfundident}
  0= \ell_2\big(\ell_3(a_1,a_2,a_3),M\big)&+\ell_3\big(\ell_2(M,a_1),a_2,a_3\big)\\
     &+\ell_3\big(\ell_2(M,a_2),a_3,a_1\big)+\ell_3\big(\ell_2(M,a_3),a_1,a_2\big)\,.
}
In the second  case, the condition ${\cal J}_4(a_1,a_2,a_3,a_4)=0$ reduces to
  \eq{
  \label{newfundidentb}
  0= \ell_2\big(\ell_3(a_1,a_2,a_3),a_4\big)&-\ell_2\big(\ell_3(a_2,a_3,a_4),a_1\big)\\
     &+\ell_2\big(\ell_3(a_3,a_4,a_1),a_2\big)-\ell_2\big(\ell_3(a_4,a_1,a_2),a_3\big)\,.
}
Since all higher relations ${\cal J}_n=0$, $n>4$ are trivially satisfied in the 2-term
truncation \eqref{linftytwoterm} we have shown  that the algebra of
bi- and cubic matrices can be extended to
a 2-term $L_\infty$ algebra, which we denote as $L^{\rm cub}_2$ in the following.

Following \cite{Hohm:2017cey}, it is an important question whether
this algebra can be extended to a three-term $L_\infty$
algebra. However, our current construction actually fits into the
no-go theorem of \cite{Hohm:2017cey} for such an extension. This is
due to the fact that $a\cdot a=M$ and therefore cubic elements do not
span an ideal.  As we will discuss at the end of section
\ref{sec:extended_gaug_theories},
it is possible to
circumvent this issue on a formal level, with the caveat that one
needs to construct a compatible inner product, which turns out to be a
formidable task.
  On that note, let us stress that the inner product \eqref{Z2innprod} is $\mathbb{Z}_2$-graded and not to be confused with the
  more special notion of a graded cyclic inner product on an
  $L_\infty$ algebra \eqref{gradedinnprod}.

\subsection{Summary of \texorpdfstring{$L^{\rm cub}_2$}{L2cub} algebras}
\label{sec_essence}

Let us  summarize the essential
calculation rules  of such $L^{\rm cub}_2$ algebras of  (hermitian) bi- and cubic matrices.
The commutators of two such objects are defined
using the three definitions of products \eqref{prodbibi},\eqref{prodcubiccubic} and \eqref{prodbicubicf}.
These are non-associative only for three cubic matrices,
which leads to the definition of a three-bracket in terms
of the respective Jacobiator, i.e.
   \eq{
            [a_1,a_2,a_3]:=-{\rm Jac}(a_1,a_2,a_3)\,.
}
Then, the commutator bracket and the so-defined three-bracket
satisfy the two fundamental identities
\eq{
  \label{newfundidenttota}
  [M,[a_1,a_2,a_3]]&=[[M,a_1],a_2,a_3]+[a_1,[M,a_2],a_3]+[a_1,a_2,[M,a_3]]\,,
 }
\eq{
  \label{newfundidenttotb}
  0&=[[a_1,a_2,a_3],a_4] - [[a_2,a_3,a_4],a_1]+[[a_3,a_4,a_1],a_2]-[[a_4,a_1,a_2],a_3]\,.
}
Another property that we will also need in the following
is the cyclicity of the trace  \eqref{traceprop}.

For a more compact notation, we will eventually join bi- and cubic
matrices into a compact object ${\cal A}=M\oplus a$.  The fundamental
identity for such  objects is then the relation \eqref{jfourtotal}
\eq{
  \label{newfundidtotal}
        [ [{\cal A}_1,{\cal A}_2,{\cal A}_3],{\cal A}_4]\ -\ &[ [{\cal
          A}_2,{\cal A}_3,{\cal A}_4],{\cal A}_1]+[ [{\cal A}_3,{\cal
          A}_4,{\cal A}_1],{\cal A}_2]-[ [{\cal A}_4,{\cal A}_1,{\cal
          A}_2],{\cal A}_3]\\
        &=[ [{\cal A}_1,{\cal A}_2],{\cal A}_3,{\cal A}_4]-[ [{\cal A}_2,{\cal
    A}_3],{\cal A}_4,{\cal A}_1]+[ [{\cal A}_3,{\cal A}_4],{\cal
    A}_1,{\cal A}_2]\\
  &\,-[ [{\cal A}_4,{\cal A}_1],{\cal A}_2,{\cal
    A}_3]-[ [{\cal A}_1,{\cal A}_3],{\cal A}_2,{\cal A}_4]-[ [{\cal
    A}_2,{\cal A}_4],{\cal A}_1,{\cal A}_3]\,.
 }

 \subsection{A class of \texorpdfstring{$L^{\rm cub}_2$}{L2cub} algebras based on  Lie algebras }
 \label{sec_SU2ex}

 In order to show that we are not talking about an empty set,
 let us first explicitly construct a simple  $L^{\rm cub}_2$ algebra
 based  on an ${\cal N}$-dimensional irreducible representation of  the $SU(2)$ Lie algebra with spin $j$.
 Hence,  we have ${\cal N}=2j+1$ and the commutation relation
\eq{
            [\lambda^i,\lambda^j]=i\, \sum_k \epsilon^{ijk} \,\lambda^k\,.
}
Now  choose the three generators in $V_B$ (bimatrices) as $T^i=\lambda^i$
and $3{\cal M}$ hermitian cubic ${\cal M}\times {\cal N}\times {\cal N}$ matrices in $V_C$ as 
\eq{
  \label{thesearethegs}
  u_r^i=(0,\ldots,0,\underbrace{\lambda^i}_{r-{\rm th}},0,\ldots,0)\,,
}
where we have expressed each cubic matrix as an ${\cal M}$-tupel  of
$ {\cal N}\times {\cal N}$ matrices.
Then one can straightforwardly determine the following commutation
relations
\eq{
  \label{su2cubiclag}
  [T^i,T^j]&=i\, \sum_k \epsilon^{ijk} \,T^k\,,\qquad  [T^i,u_r^j]=i\,
  \sum_k \epsilon^{ijk} \,u_r^k\,,\qquad [u_r^i,u_s^j]=i\delta_{rs}\,
  \sum_k \epsilon^{ijk} \,T^k \,.
}
Moreover,  there exist non-vanishing three-brackets
\eq{
  \label{sutwothreebr}
  [u^i_r,u_r^j,u_s^k]= \sum_l (\delta^{ik}\delta^{jl}-\delta^{il}\delta^{jk})
  \,u_s^l\,,\qquad r\ne s\,.
 }
 Hence, for each value of ${\cal M}$ and ${\cal N}$, these three  bimatrices and
 $3{\cal M}$  cubic matrices
 can be regarded as a  representation of this non-associative
 $L^{\rm cub}_2$ algebra.
 
This is just a simple example and one can easily generalize
it by replacing $SU(2)$ by a more general Lie algebra $G$.
One only has to adapt  the structure constant in the commutation relations
 \eqref{su2cubiclag} to $f^{ijk}$, 
in which case the three-bracket  becomes e.g.
\eq{
  [u_r^i,u_r^j,u_s^k]= \sum_{l,m} f^{ijm}  f^{mkl} \,u_s^l\,,\qquad r\ne s\,.
}
This defines a large class of such $L^{\rm cub}_2$ algebras but
there certainly exist many other    $L^{\rm cub}_2$ algebras.
Their further study or even classification is not the topic of
this paper, but might deserve a deeper mathematical investigation.

\subsection{A comment on Lie 3-algebras}

Finally, let us make a  comment about the relation of these
algebras to so-called Lie  3-algebras or Filippov 3-algebras,
which appeared in the BGL-theory for multiple
membranes \cite{Bagger:2006sk,Gustavsson:2007vu,Bagger:2007jr}.
The latter algebras are the quantum versions of Nambu-brackets and 
are defined by a multi-linear 3-bracket
\eq{
  [\cdot,\cdot,\cdot]:V\otimes V\otimes V\to V
  }
  satisfying total antisymmetry and the fundamental identity
  \eq{
    \label{fundidentnambu}
    [A_1,A_2,[A_3,A_4,A_5]]=
    &[[A_1,A_2,A_3],A_4,A_5]+[A_3,[A_1,A_2,A_4],A_5]\\
    &+[A_3,A_4,[A_1,A_2,A_5]]\,.
  }
Clearly, for these algebras there is a priori no accompanying 2-bracket 
and the fundamental identity involves two 3-brackets instead of
one 2-bracket and one 3-bracket, as  in our case.
Therefore, generically the 3-brackets are not expected to satisfy
the  fundamental identity \eqref{fundidentnambu}, as it 
does not follow directly from the definition of the three-bracket
as a Jacobiator.

First, let us observe that  after renaming e.g. $y^1=u_r^1$,  $y^2=u_r^2$,  $y^3=u_s^1$,  $y^4=u_s^2$
with $r\ne s$,  the three-brackets  for these four cubic matrices satisfy
 \eq{
            [y^i,y^j,y^k]= \epsilon^{ijkl} \,y^l\,,
} 
which is a quantum version of the Nambu-bracket on $S^3$
and hence satisfies also the fundamental identity
\eqref{fundidentnambu}.
Forgetting the intermediate bi-matrices, we could now
directly define a three-product of three cubic matrices
by the associator, which in components reads
\eq{
  \label{ternbracket}
         (a\cdot b\cdot c)_{rij}:=\sum_{s,m,n} \big(a_{rim} \, c_{snj}-
         a_{sim}\,  c_{rnj}\big)  b_{smn}\,.
}
Note that this   definition  is different from the ones proposed
in \cite{Awata:1999dz,Kawamura:2002yz,Kawamura:2003cw}.
Then the three-bracket can be defined as
\eq{
      [a,b,c]=a\cdot b\cdot c+b\cdot c\cdot a+c\cdot a\cdot b-a\cdot c\cdot b
               -b\cdot a\cdot c-c\cdot b\cdot a\,.
             }
Next, the question is whether this generalizes  to 
the full  initial $G=SU(2)$ example.
That this is not case is revealed by the following choice for $r\ne
s\ne t\ne r$
\eq{
        a_1=u_t^2\,,\quad  a_2=u_s^3\,,\quad  a_3=u_r^1\,,\quad
        a_4=u_r^2\,,
        \quad  a_5=u_s^1\,,
  }      
  for which the left  hand side of \eqref{fundidentnambu} is equal
  to $u_t^3$ and the right  hand side vanishes.

  Let us conclude  that, as already proposed in \cite{Awata:1999dz,Kawamura:2002yz,Kawamura:2003cw},
  cubic matrices can realize Lie 3-algebra structures,
  where the intermediate step of first constructing
  $L^{\rm cub}_2$ algebras of bi- and cubic matrices seems
  to allow for   a more systematic approach.
  Following this further is beyond the scope
  of this paper.

\section{Gauge theories for  \texorpdfstring{$L^{\rm cub}_2$}{L∞} algebras }

An important question is whether such a non-associative structure
is compatible with physics, i.e. whether one can extend
physical theories involving bimatrices to similar theories
involving the $L^{\rm cub}_2$ structure. 
Here, we consider  a potential generalization of Yang-Mills theory.
As mentioned in the introduction, one motivation is to find
a generalization of the BFSS matrix model and the latter
can indeed be considered as the dimensional reduction
of Yang-Mills theory from ten to one dimension, i.e. on the
world-line of $D0$-branes. Hence, this seems
to be a reasonable first approach. 

In this section, we sort of bootstrap the structure of $L^{\rm cub}_2$
gauge theories in a step by step procedure.
In this manner we essentially rederive the relations
found for a general abstract  two-term $L_\infty$ algebra in \cite{Zucchini:2011aa}
(see also \cite{Ritter:2013wpa,Jurco:2014mva}).

\subsection{Yang-Mills theory in a nutshell}

Recall that for a non-abelian Yang Mills theory, the gauge potential
$A_\mu$ takes values in the adjoint representation of a Lie algebra 
\eq{
        A_\mu=\sum_i  A_\mu^i\, T_i
}
and as such is  matrix valued.
Its infinitesimal gauge variation reads
\eq{
    \delta_{\Lambda} A_\mu=\partial_\mu \Lambda +i [\Lambda,A_\mu]
  }
so that the field strength
\eq{
              F_{\mu\nu}=\partial_\mu A_\nu- \partial_\nu A_\mu-i [A_\mu,A_\nu]    
}
transforms covariantly as
\eq{
              \delta_{\Lambda}  F_{\mu\nu}=i [\Lambda,F_{\mu\nu}] \,. 
            }
Note that in order to show this, one invokes the Jacobi identity.
Furthermore one introduces a covariant derivative $\mathcal{D}_\mu$, the action of which on a matter field $\psi$ in the adjoint representation is given by
\begin{eqnarray}
    D_{\mu} \psi = \partial_\mu \psi - i[A_{\mu},\psi]\,.
\end{eqnarray}
It is then immediate that we can express $F_{\mu\nu}$ via
\begin{equation}
    [D_{\mu},D_{\nu}]=-i F_{\mu\nu}\,.
\end{equation}
Invoking also the  cyclicity of the trace, the action
\eq{
            S=-{\frac{1}{4}}\int d^D x\; {\rm tr}\Big(F_{\mu\nu}\, F^{\mu\nu} \Big)
}
is gauge invariant and leads to the equation of motion
\eq{
  \label{eomordym}
  D_{\nu}F^{\mu\nu}=0\,.
 }   
In addition, one can explicitly check the trivial Bianchi identity\footnote{As usual $n$ indices in brackets
  are completely antisymmetrized with a factor $1/n!$ in front.}
\eq{
  \label{ymbianchi}
          3\,D_{[\mu} F_{\nu\rho]}=    D_{\mu} F_{\nu\rho}+D_{\nu} F_{\rho\mu}+D_{\rho} F_{\mu\nu}=0\,.
 }
 Let us now try to generalize this construction to our case,
 where we will see that we are forced to introduce also higher form
 gauge fields.
 
\subsection{One-form gauge theories for  \texorpdfstring{$L^{\rm cub}_2$}{L2cub}}

Instead of a one-form gauge field taking values in a Lie algebra
we now consider the case where  it takes values
in an  $L^{\rm cub}_2$ algebra. 
This means that the total gauge field ${\cal A}_\mu$ is expanded into both bimatrices and
cubic matrices
\eq{
  \label{totalfieldexpand}
           {\cal A}_\mu=A_\mu + a_\mu =\sum_i  A^i_\mu \, T_i +\sum_r  a^r_\mu \,u_r \in V_B \oplus  V_C\,,
}
where in the following computation the explicit form of the bimatrices $T_i$ and
the cubic matrices $u_r$ is never needed. Analogous to  $A_\mu$ and $a_\mu$,
in the following capital letters denote   bimatrices and small
letters cubic matrices.
Moreover, for pedagogical reasons we present the computation for
``component'' fields, like   $A_\mu$ and $a_\mu$, and only in the very
end write the result in terms of the total field  ${\cal A}_\mu$.

We start with  the gauge variation and  write down the most general commutators
consistent with the $\mathbb Z_2$ grading
\eq{
  \delta_{(\Lambda,\lambda)} \,  A_\mu&=\partial_\mu \Lambda +i [\Lambda,A_\mu]+i [\lambda,a_\mu]\,,\quad\qquad 
  \delta_{(\Lambda,\lambda)} \,  a_\mu=\partial_\mu \lambda +i [\Lambda,a_\mu]+i [\lambda,A_\mu]\,.
  }
Similarly, for the field strength we define
\eq{
       F_{\mu\nu}&=\partial_\mu A_\nu- \partial_\nu A_\mu-i [A_\mu,A_\nu]-i [a_\mu,a_\nu]\,, \\[0.1cm]     
       f_{\mu\nu}&=\partial_\mu a_\nu- \partial_\nu a_\mu-i [a_\mu,A_\nu]-i [A_\mu,a_\nu]    \,,   
}  
which under a gauge variation transform as
\eq{
  \delta_{(\Lambda,\lambda)} \,   F_{\mu\nu}&=i [\Lambda,F_{\mu\nu}] +i [\lambda,f_{\mu\nu}] \,,\\[0.1cm]
   \delta_{(\Lambda,\lambda)} \,   f_{\mu\nu}&=i [\Lambda,f_{\mu\nu}] +i [\lambda,F_{\mu\nu}]-[\lambda,a_\mu,a_\nu]\,.
 }
 The last term in the second row is anomalous, in the sense that it breaks
 the gauge covariance and can be traced back
 to the non-vanishing Jacobiator for three cubic matrices.
 However, one can repair this failure by introducing a pair of   two-forms,
 redefine the field strength
 \eq{
             \hat F_{\mu\nu}=F_{\mu\nu}+B_{\mu\nu}\,,\qquad \hat f_{\mu\nu}=f_{\mu\nu}+b_{\mu\nu}
 }
and impose the following gauge variations
\eq{
        \delta_{(\Lambda,\lambda)} \,  B_{\mu\nu}=i[\Lambda,B_{\mu\nu}] +i[\lambda,b_{\mu\nu}]\,,\qquad 
        \delta_{(\Lambda,\lambda)} \,  b_{\mu\nu}=i[\Lambda,b_{\mu\nu}] +i[\lambda,B_{\mu\nu}] + [\lambda,a_\mu,a_\nu]\,.
}
Then, one gets
\eq{
  \delta_{(\Lambda,\lambda)} \,   \hat F_{\mu\nu}&=i [\Lambda,\hat F_{\mu\nu}] +i [\lambda,\hat f_{\mu\nu}] \,,\qquad 
   \delta_{(\Lambda,\lambda)} \,   \hat f_{\mu\nu}=i [\Lambda,\hat f_{\mu\nu}] +i [\lambda,\hat F_{\mu\nu}]\,.
 }
 These two-forms come with their own gauge symmetries acting as
 \eq{
   \label{bfieldvaria}
            \delta_{(\Xi,\xi)} \,  B_{\mu\nu}&=2\Big(\partial_{[\mu} \Xi_{\nu]}  -i \big[\Xi_{[\mu},A_{\nu]}\big]-i \big[\xi_{[\mu},a_{\nu]}\big]\Big)\,,\qquad  \delta_{(\Xi,\xi)} \, A_{\mu}=-\Xi_\mu\,, \\[0.1cm]
            \delta_{(\Xi,\xi)} \, b_{\mu\nu}&=2\Big(\partial_{[\mu} \xi_{\nu]} -i \big[\xi_{[\mu},A_{\nu]}\big] -i \big[\Xi_{[\mu},a_{\nu]}\big]\Big)\,,\qquad\phantom{a}  \delta_{(\Xi,\xi)} \, a_{\mu}=-\xi_\mu\,, 
} 
so that both $\hat F_{\mu\nu}$ and $\hat f_{\mu\nu}$ are invariant, i.e.
$ \delta_{(\Xi,\xi)} \, \hat F_{\mu\nu}= \delta_{(\Xi,\xi)} \, \hat f_{\mu\nu}=0$.

With these ingredients and the cyclicity property \eqref{traceprop} of  the trace,
 it is now straightforward to  show that the action
 \eq{
   \label{actionfirstlevel}
  S=-{\frac{1}{4}}\int d^D x\; \Big( {\rm tr}\big(\hat F_{\mu\nu}\, \hat F^{\mu\nu} \big)
  +{\rm tr}\big(\hat f_{\mu\nu}\, \hat f^{\mu\nu} \big)\Big)
}
is invariant under all kinds of gauge transformations $(\Lambda,\lambda,\Xi_\mu,\xi_\nu)$.
In this action the one-form gauge fields $A_\mu$, $a_\mu$ are 
dynamical and the two-form fields  $B_{\mu\nu}$, $b_{\mu\nu}$ do not
have a kinetic term yet.
Using again the cyclicity of the trace \eqref{traceprop}, the 
resulting equations of motion
for $A_\mu$ and $a_\mu$ read
\eq{
  \partial_\nu \hat F^{\mu\nu}+i [\hat F^{\mu\nu}, A_\nu]+i [\hat f^{\mu\nu}, a_\nu]&=0\,,\\[0.1cm]
  \partial_\nu \hat f^{\mu\nu}+i [\hat f^{\mu\nu}, A_\nu]+i [\hat F^{\mu\nu}, a_\nu]&=0\,.
}
Variation of the action with respect to the two-form fields yields
the constraints
\eq{
  \label{bbvaria}
  \hat F_{\mu\nu}=\hat f_{\mu\nu}=0\,.
 } 
In addition one can show the following two Bianchi-identities
\eq{
  \label{Bianchiff}
  &3\Big(\partial_{[\mu} \hat F_{\nu\rho]} + i \big[\hat F_{[\mu\nu},A_{\rho]}\big] + i \big[\hat f_{[\mu\nu},a_{\rho]}\big]  \Big) =H_{\mu\nu\rho}\,,\\[0.1cm]
   &3\Big(\partial_{[\mu} \hat f_{\nu\rho]} + i \big[\hat f_{[\mu\nu},A_{\rho]}\big] + i \big[\hat F_{[\mu\nu},a_{\rho]}\big]\Big)  =h_{\mu\nu\rho}\,,
 }
 with the three forms
 \eq{
   \label{threeformstrength}
   H_{\mu\nu\rho}&=3\Big(\partial_{[\mu} B_{\nu\rho]} +i \big[B_{[\mu\nu},A_{\rho]}\big]+i \big[b_{[\mu\nu},a_{\rho]}\big]\Big)\,,\\[0.1cm]
   h_{\mu\nu\rho}&=3\Big(\partial_{[\mu} b_{\nu\rho]} +i \big[b_{[\mu\nu},A_{\rho]}\big]+i \big[B_{[\mu\nu},a_{\rho]}\big]\Big)-[a_\mu,a_\nu,a_\rho]  \,,
 }
 which seem to be natural generalizations of the three-form field strengths of the
 two-form gauge fields.
Note, in particular, the appearance of the three-bracket in the second row,
which only depends on the one-form gauge field $a_\mu$.

Hence, we have shown that, up to this point, one can consistently formulate
a gauge theory  based on  a non-associative $L^{\rm cub}_2$ algebra.
However, we have seen that the formalism forces us 
to introduce additional  two-form gauge fields which transform non-trivially under
a zero-form gauge transformation. In addition, the two-forms
have their own one-form gauge symmetry under which also
the gauge fields $(A_\mu,a_\mu)$ transform non-trivially.
Hence, in the next step one would like to include kinetic terms
for these fields, as well. However, before doing this
let us write the relations from this section in a more concise form.

\subsection{Compact version of two-form extended gauge theory  for  \texorpdfstring{$L^{\rm cub}_2$}{TEXT}}\label{subsec_33}
    
 Using now the total fields, like ${\cal A}_\mu$ from \eqref{totalfieldexpand},
 let us formalize  our results for a  one-form  gauge theory with
 background fields.
 First, we consider fields $\mathit{\Lambda}, {\cal A}_\mu$
 as elements of $\Omega^*(M)\otimes V_0$, where
 $V_0$ is the degree zero vector space of the $L_\infty$ algebra
 and $M$ denotes $D$-dimensional Minkowski space.
 The  infinitesimal gauge variation of the gauge potential
 reads
\eq{
    \delta_{\mathit{\Lambda}} {\cal A}_\mu=\partial_\mu \mathit{\Lambda} -i
    [{\cal A}_\mu,\mathit{\Lambda}]= \mathcal{D}_{\mu} \mathit{\Lambda}\,,
  }
where we introduced a ``covariant'' derivative\footnote{This
   derivative is not really covariant, as one finds
   $[\mathcal{D}_{\mu},\mathcal{D}_{\nu}]=-i\mathcal{F}_{\mu\nu}$ and not $\hat{\mathcal{F}}_{\mu\nu}$.}.
  The corresponding field strength
\eq{
             {\cal  F}_{\mu\nu}=\partial_\mu {\cal A}_\nu-
             \partial_\nu {\cal A}_\mu-i [{\cal A}_\mu,{\cal A}_\nu]    
}
transforms as
\eq{
              \delta_{\mathit{\Lambda}}  {\cal F}_{\mu\nu}=i [\mathit{\Lambda},{\cal F}_{\mu\nu}]
              -\ell_1\big([\mathit{\Lambda},{\cal A}_\mu,{\cal A}_\nu]\big)\,, 
            }
where we have explicitly written the map $\ell_1$ so
that   ${\cal F}_{\mu\nu}\in\Omega^2(M)\otimes V_0$.
In the following, we will  not explicitly include $\ell_1$ in the
relations,
it is understood that it is implicitly present in front of  every three-bracket.
            
To make the field strength  gauge covariant we added  a background 2-form field
${\cal B}_{\mu\nu}\in \Omega^2(M)\otimes V_0$  transforming as
\eq{
              \delta_{\mathit{\Lambda}}  {\cal B}_{\mu\nu}=i [\mathit{\Lambda},{\cal B}_{\mu\nu}]
              +[\mathit{\Lambda},{\cal A}_\mu,{\cal A}_\nu]\,. 
            }
There exists also  a one-form gauge symmetry under which the gauge
fields transform
as
\eq{
            \delta_{\mathit{\Xi}}  {\cal B}_{\mu\nu}=2\,
            \mathcal{D}_{[\mu}  \mathit{\Xi}_{\nu ]} \,,\qquad\qquad 
            \delta_{\mathit{\Xi}}  {\cal A}_{\mu} =- \mathit{\Xi}_\mu \,.
          }
The improved field strength $\hat{\cal F}_{\mu\nu}= {\cal
  F}_{\mu\nu}+{\cal B}_{\mu\nu}$ satisfies the Bianchi identity
\eq{
  \label{bianchifftotal}
                3\,\mathcal{D}_{[\mu} \hat{\cal F}_{\nu\rho]}= {\cal H}_{\mu\nu\rho}\,,
 }
with  the three-form field strength
 \eq{
   \label{bianchiFcompact}
                {\cal H}_{\mu\nu\rho}=3\,\mathcal{D}_{[\mu}\mathcal{B}_{\nu\rho]}
                -[ {\cal A}_{\mu}, {\cal A}_{\nu}, {\cal A}_{\rho}]\,.
              }
Then, the action
\eq{
  \label{compactactionF}
            S= -{\frac{1}{4}}\int d^{D} x\;{\rm tr}\big(\hat{\cal
              F}_{\mu\nu}\, \hat{\cal F}^{\mu\nu} \big)
}
is gauge invariant and  leads to the equations of motion
\eq{
      \mathcal{D}_\nu \hat{\cal F}^{\mu\nu} =0\,,\qquad\qquad  \hat{\cal F}_{\mu\nu} =0\,,
    }
    where the second, topological one follows from the variation with respect
    to the non-dynamical field ${\cal B}_{\mu\nu}$.
One can show that the gauge variation of this equation of motion
is non-canonical
\eq{
      \delta_{\mathit\Lambda} \Big(\mathcal{D}_\nu \hat{\cal
        F}^{\mu\nu}\Big)=i[\mathit\Lambda,\mathcal{D}_\nu \hat{\cal
        F}^{\mu\nu}]+ [\mathit\Lambda, {\cal A}_\nu,\hat{\cal F}^{\mu\nu}]\,
 }       
and only reduces to  the familiar form upon invoking the equation of motion for the non-dynamical field ${\cal B}_{\mu\nu}$.

Hence in this compact notation many of the relations take almost the same
form as for ordinary Yang-Mills theories, though with the essential
differences that the non-associativity reveals itself
by the appearance of the three-bracket, that one needs the two-form field compensating for a
resulting non-standard gauge-transformation behavior and that
the equations of motion are of topological type.

\subsection{Three-form extended gauge theories for  \texorpdfstring{$L^{\rm cub}_2$}{TEXT}}\label{sec:extended_gaug_theories}

The Bianchi identity for the gauge field ${\cal A}_\mu$ has revealed
an expression \eqref{bianchiFcompact}
for the generalized field strength ${\cal H}_{\mu\nu\rho}\in
\Omega^3(M)\otimes V_0$ of  the gauge field ${\cal B}_{\mu\nu}$.
Hence, the next question is whether the action can be extended
by also making these fields dynamical.
By varying the expression \eqref{bianchiFcompact} for the three-form
field strength and after employing the fundamental identity \eqref{newfundidtotal}, we find
\eq{
  \label{Hhgaugevarai}
  \delta_{\mathit\Lambda} \,  {\cal H}_{\mu\nu\rho}&=i \big[\mathit\Lambda,{\cal H}_{\mu\nu\rho}\big]+
  3 \,\big[\mathit\Lambda,\hat{\cal F}_{[\mu\nu},{\cal A}_{\rho]}\big]\,.
 }
 This is analogous to the  variation of the previous two-form field
 strength ${\cal F}_{\mu\nu}$ including an  unconventional  three-bracket term.
 Next we compute the variation under the one-form gauge variation
 $\mathit{\Xi}_\mu$, which takes
 the form
 \eq{
   \label{bfieldvariaH}
   \delta_{\mathit{\Xi}} \,  {\cal H}_{\mu\nu\rho}&=3i\, \big[\mathit{\Xi}_{[\mu},\hat{\cal F}_{\nu\rho]}\big]\,.   
 } 
 Note that the three-bracket term in the definition of ${\cal H}_{\mu\nu\rho}$ \eqref{bianchiFcompact}
 cancels against a non-vanishing Jacobiator. Now, we could proceed in
 two ways.

 \paragraph{Gauge rectifier:} First, one could try to repair
 the anomalous gauge transformation behavior by introducing
 a so-called gauge rectifier by redefining
 \eq{
          \hat{\cal H}_{\mu\nu\rho}= {\cal H}_{\mu\nu\rho}+\Delta({\cal
            F},{\cal A})\,.
 } 
This concept  was introduced
in \cite{Zucchini:2011aa} to make the gauge field ${\cal H}_{\mu\nu\rho}$ covariant
without introducing a new three-form gauge field  ${\cal C}_{\mu\nu\rho}$.
Then, one could add also
the three-form field strength to the action
\eq{\label{twoformgaugeaction}
  S=\int d^D x\; \bigg(&-{\frac{1}{4}} {\rm tr}\big(\hat{\cal
    F}_{\mu\nu}\, \hat{\cal F}^{\mu\nu} \big)-{\frac{1}{6}}\;  {\rm
    tr}\big(\hat{\cal H}_{\mu\nu\rho}\, \hat{\cal H}^{\mu\nu\rho}  \big)\bigg)\,,
}
where the relative normalization has been chosen such that one
gets standard kinetic terms.
By construction, this action  is invariant under the two kinds of
gauge transformations $(\mathit{\Lambda},{\mathit\Xi}_\mu)$.
Using again the cyclicity of the trace \eqref{traceprop}, the  equations of motion
resulting from  the variation $\delta \mathcal{A}_\mu$  read
\eq{
   \label{eomh96a}
  \mathcal{D}_\nu \hat{\cal F}^{\mu\nu}
  +i [\hat{\cal H}^{\mu\nu\rho}, {\cal B}_{\nu\rho}]-[\hat{\cal
    H}^{\mu\nu\rho},{\cal A}_\nu,{\cal A}_\rho]&=0\,,
}
where the last three-bracket term
follows  from the variation of the $[{\cal A}_\mu,{\cal A}_\nu,{\cal
  A}_\rho]$ term in the definition of
$\hat{\cal  H}^{\mu\nu\rho}$.
From the variation $\delta {\cal B}_{\mu\nu}$  we analogously obtain
the equation of motion
\eq{
   \label{eomh96b}
  \mathcal{D}_\rho \hat{\cal H}^{\mu\nu\rho} -{\frac{1}{2}} \hat{\cal
    F}^{\mu\nu}=0\,.
}
This was the general story, however in our concrete case, the only gauge rectifier we found implies the redefinition
\eq{
  \label{eomh96c}
          \hat{\cal H}_{\mu\nu\rho}={\cal H}_{\mu\nu\rho} - 3 \mathcal{D}_{[\mu}
          \hat{\cal F}_{\nu\rho]} =0\,,
        }
which identically vanishes due to the Bianchi-identity. 
Hence, using this extra condition, solving the equations
of motion \eqref{eomh96a} and  \eqref{eomh96b} 
leads to
$\hat{\cal  F}_{\mu\nu}=0$ and we are effectively back
to the theory from the previous subsection \ref{subsec_33}.

The appearance of this fake-flatness condition  was already observed in
\cite{Saemann:2019leg}.
Solving the equation 
$\hat{\cal  F}_{\mu\nu}=0$ allows to express ${\cal B}_{\mu\nu}$
in terms of the gauge field ${\cal A}_\mu$ without
any extra dynamical constraint on ${\cal A}_\mu$.
Moreover, it means  that the equations of motion \eqref{eomh96a}-\eqref{eomh96c}
do not reduce to the one \eqref{eomordym} of usual Yang-Mills theory upon setting
all cubic fields and all higher form fields to zero. Indeed, 
the equation of motion \eqref{eomh96b} still implies the flatness
condition  $F_{\mu\nu}=0$,
which is rather the equation of motion of Chern-Simons theory in 3D
than of Yang-Mills theory.

To summarize, we have arrived at the theory generally proposed in
\cite{Zucchini:2011aa}, though now with a concrete realization
of the underlying 2-term $L_\infty$ algebra.
Indeed, in this case we could also consider the objects
$\mathit{\Lambda}, {\cal A}_\mu, \hat{\cal F}_{\mu\nu}$ as elements
in $\Omega^*(M,V_0)$ and $\mathit{\Xi}_\mu, {\cal B}_{\mu\nu}, \hat{\cal H}_{\mu\nu\rho}$
as elements in $\Omega^*(M,V_1)$. For consistency, we would then
define
$\hat{\mathcal{F}}_{\mu\nu}=\mathcal{F}_{\mu\nu}+\ell_1(\mathcal{B}_{\mu\nu})$
and modify the gauge tranformation of $\mathcal{A}_\mu$ under
$\mathit{\Xi}_\mu$ to
$\delta_{\mathit{\Xi}}\mathcal{A}_\mu=-\ell_1(\mathit{\Xi}_\mu)$.
Up to such $\ell_1$ insertions, all relations  would be unaffected.

\paragraph{Background 3-form:}  A second possibility to   repair
the non-associative anomaly in \eqref{Hhgaugevarai}
is to leave all objects in  $\Omega^*(M,V_0)$ and
introduce a  new  background three-form ${\cal C}_{\mu\nu\rho}\in
\Omega^*(M,V_0)$ which  redefines the field strength as 
 \eq{
             \hat{\cal H}_{\mu\nu\rho}={\cal H}_{\mu\nu\rho}+{\cal C}_{\mu\nu\rho}\,.
 }
Imposing the following zero- and one-form gauge variations
\eq{  
        \delta_{\mathit{\Lambda}} \,  {\cal C}_{\mu\nu\rho}=i[\mathit{\Lambda},{\cal C}_{\mu\nu\rho}] 
        -3\big[\mathit{\Lambda},\hat{\cal  F}_{[\mu\nu},{\cal A}_{\rho
          ]}\big]\,,\qquad\quad
          \delta_{\mathit{\Xi}} \,  {\cal C}_{\mu\nu\rho}=-3i\, \big[\mathit{\Xi}_{[\mu},\hat{\cal F}_{\nu\rho]}\big] \,,
        }
the three-form $\hat{\cal H}_{\mu\nu\rho}$ transforms
covariantly under the gauge variation $\delta_\mathit{\Lambda}$
and is invariant under $\delta_{\mathit{\Xi}}$.
In this case, we would get the same equations of motion
\eqref{eomh96a},\eqref{eomh96b} with the former constraint
\eqref{eomh96c} now resulting as the equation of motion
for  ${\cal C}_{\mu\nu\rho}$.

However, there is an issue arising, as
we can now impose
also  a new 2-form gauge variation of the gauge fields via
\eq{
  \delta_{\mathit\Theta} \,  {\cal C}_{\mu\nu\rho}=& 3\,{\cal
    D}_{[\mu} {\mathit\Theta}_{\nu\rho]}\,,
  \qquad  \delta_{\mathit\Theta} \,  {\cal
    B}_{\mu\nu}=-{\mathit\Theta}_{\mu\nu}\,,
  \qquad
    \delta_{\mathit\Theta} \,  {\cal
    A}_{\mu}=0\,.
}    
With these assertions the three form  $\hat{\cal
  H}_{\mu\nu\rho}$ is
invariant under the  two-form gauge variation, but the
two-form field strength transforms as
$\delta_{\mathit\Theta} \hat{\cal F}_{\mu\nu}=-{\mathit\Theta}_{\mu\nu}$
so that its kinetic term in the action is not gauge invariant. 
For this reason, we are not pursuing this direction
further.

Let us comment that working with a more general $L_\infty$ algebra
with more non-vanishing vector spaces, 
one can indeed continue constructing higher form gauge fields
and gauge invariant field strength.
This seems to be 
reminiscent of the tensor hierarchy appearing
 in  gauged maximal supergravity \cite{deWit:2005hv,deWit:2008ta,deWit:2008gc}
 and  Exceptional Field Theory \cite{Aldazabal:2013via,Hohm:2015xna}.
 However, as shown in \cite{Bonezzi:2019ygf}, based on  $L_\infty$ algebras
 one solely arrives at a topological tensor hierarchy.
 For really getting the dynamical tensor hierarchy of gauged maximal
 supergravity,
 the relevant structure is an infinity-enhanced Leibniz(-Loday) algebra \cite{Lavau:2017tvi}.
 Hence, it is an interesting question whether cubic
 or even higher index matrices, like the four index object from
 eq. \eqref{tensorfourindex},
 can be used to provide
 concrete examples of such Leibniz algebras.

\paragraph{Extension to an $L_3^{\rm cub}$:} Coming back to the
discussion at the end of section \ref{sec_cubic}, let us analyze
whether one can evade the no-go of \cite{Hohm:2017cey} and get an
extension to at least an $L_3^{\rm
  cub}$ algebra. For that purpose we need  to modify the  initial products so that the cubic matrices can form an ideal.
Hence, one needs a product of cubic matrices that gives again a cubic
matrix and in turn use this multiplication in order to define
$\ell_2$.  An obvious candidate is 
\begin{align}
 a \star b := (a \cdot b)\cdot \mathbb{1}^c = \Big(\sum_{r=1}^\mathcal{M}
  a_r \,b_r\Big) \cdot\mathbb{1}^c\,\qquad {\rm with}\quad 
    \mathbb{1}^c = (\mathbb{1},\dots,\mathbb{1})\,.
\end{align}
We would no longer have a  $\mathbb Z_2$ grading but
\begin{equation}
     M\cdot M \to M\,,\qquad  a \star  a\to a\,,\qquad  M\cdot  a \to a\,,\qquad  a\cdot  M \to a\,.
\end{equation}

To properly realize such an extension we would need to recheck all relations $\mathcal{J}_i=0$. It turns out that this is possible by defining the maps $\ell_i$ in the following way (where $\mathcal{A}_i \in \mathcal{V}^0=V^0_B \oplus V^0_C,\, \mathcal{B}_i \in \mathcal{V}^1=V^1_B \oplus V^1_C$ and $c_i \in \mathcal{V}^2=V_C$)
\begin{align}
\begin{aligned}
    \ell_2(\mathcal{A}_1,\mathcal{A}_2)&=[\mathcal{A}_1,\mathcal{A}_2] = [A_1,A_2] \in V_B^0 + ([A_1,a_2]+[a_1,A_2]+[a_1,a_2]_{\star}) \in V_C^0\\[0.1cm]
    \ell_2(\mathcal{A},\mathcal{B})&=([A,b] + [a,b]_\star ) \in V_C^1\\[0.1cm]
    \ell_2(\mathcal{B}_1,\mathcal{B}_2) &=\ell_2(\mathcal{C}_1,\mathcal{C}_2) = \ell_2(\mathcal{A},\mathcal{C})=\ell_2(\mathcal{B},\mathcal{C})= 0\\[0.1cm]
    \ell_3(\mathcal{A}_1,\mathcal{A}_2,\mathcal{A}_3)&={\rm Jac}(\mathcal{A}_1,\mathcal{A}_2,\mathcal{A}_3)\, \quad \text{and}\, \quad \ell_3(\mathcal{X}_1,\mathcal{X}_2,\mathcal{X}_3)=0\, \quad \text{else}\,,\\
\end{aligned}
\end{align}
where commutators involving cubic matrices are now realized via the new product.
This is conveniently summarized in the following diagram \eqref{eq:diagram}, where $\ell_2$ takes one element from the entry it starts from and one from the one the arrow ends and maps it to the latter subspace. One can uniquely determine all arguments of the $\ell_i$ by the arrows in the diagram and their degree. This includes the so-far undefined nilpotent map $\ell_1$. Note that for our purposes it will not be necessary to define a non trivial $\ell_4$ map.
\begin{equation}\label{eq:diagram}
\begin{tikzcd}[row sep=normal, column sep=normal]
  & 0   &   &  \\  
0  &  \arrow[l, "\ell_1",swap] V^0_B \arrow[rd, dashed,blue,"\ell_2"]\arrow[d, dashed,blue,"\ell_2"] \arrow[loop above,looseness=7, dashed,blue,"\ell_2"]  & \arrow[lu,"\ell_1",swap] V^1_B  &  0 \\  
0  &  \arrow[l,"\ell_1",swap] V^0_C \arrow[loop below,looseness=7, dashed,blue,"\ell_2"] \arrow[r, dashed,blue,bend right,swap,"\ell_2"] \arrow[r, dotted,red,bend right = 70,swap,"\ell_3"]  &  \arrow[l,"\ell_1",swap] V^1_C  & \arrow[lu,"\ell_1",swap]V^2_C\,
\end{tikzcd}
\end{equation}

 Observe that one can repeat the procedure of section \ref{sec_SU2ex}
 to realize such an $L_3^{\rm cub}$ using ordinary Lie
 algebras. Another immediate observation is that now that we have one
 more vector space at our disposal we have more options  in arranging
 the various gauge fields. 
 However, we will not
explore this direction further, as we were not able to construct an
inner product compatible with this structure, which means we cannot
concretely write down the corresponding
action. This approach could, nevertheless, be
useful when working on the level of the equations of motion,
as was done for $L_{\infty}$ algebras in \cite{Bonezzi:2019ygf}.

\subsection{Topological BF-theory and a deformed IKKT matrix model}

In this section, we would like to point out two other models
for  $L^{\rm cub}_2$ algebras. Since these have been more generally
discussed
in the context of abstract $L_2$ algebras\cite{Ritter:2013wpa,Ritter:2015ymv}, we keep the presentation
rather short. Let us, however, observe that our inner product, despite not being an $L_{\infty}$ graded cyclic inner product can give rise to a consistent action.

We have already seen that the proposed $L^{\rm cub}_2$ Yang-Mills theory actually
gave rise to equations of motion rather resembling those
of a topological Chern-Simons-like theory.
It is thus compelling to note that there indeed
exists an  $L^{\rm cub}_2$  generalization of the four-dimensional topological
BF-theory.
The action of this theory is
\eq{
  S_{\rm BF}\sim  \int d^4x \, \epsilon^{\mu\nu\rho\sigma}\,
      {\rm tr}\Big( {\cal B}_{\mu\nu} \big(\hat{\cal F}_{\rho\sigma}-{\frac1 2}
          {\cal B}_{\rho\sigma}\big) -{\frac 1 6} {\cal A}_\mu [{\cal A}_\nu,{\cal A}_\rho,{\cal A}_\sigma]\Big)\,.
        }
Using  the relation       
\eq{
             {\rm tr}\big( a\, [b,c,d] \big)=- {\rm tr}\big( b\,
             [c,d,a] \big)
}             
for four cubic matrices
and \eqref{newfundidtotal}  one can show that this action
is invariant under infinitesimal gauge variations $\delta_{\mathit{\Lambda}}$ and $\delta_{\mathit{\Xi}}$
up to total derivatives.
Then the variation of the action with respect to ${\cal B}_{\mu\nu}$
and ${\cal A}_\mu$
yield the two fake-flatness conditions
\eq{
  \hat{\cal F}_{\mu\nu}=0\,,\qquad
    {\cal H}_{\mu\nu\rho}=0\,.
}

As a second application we consider the dimensional reduction of
higher dimensional theories to zero dimensions, reminiscent
of the IKKT matrix models. 
In this case, we get  bi-/cubic matrix valued
matrices ${\cal X}^i$ whose dimensionally reduced gauge
transformations are
\eq{
    \delta_{\mathit\Lambda} {\cal X}^i = i[\mathit\Lambda,{\cal
      X}^i]\,,\qquad\qquad
    \delta_{\mathit\Xi} {\cal X}^i =-\mathit\Xi^i\,,
 }   
 which in particular means that they transform covariantly under $ \delta_{\mathit\Lambda} $.
 The gauge invariant field strength then becomes
 \eq{
           \hat{\cal F}^{ij}= -i[ {\cal X}^i, {\cal X}^j] +  {\cal Y}^{ij}\,,
}
where  the two-index object $ {\cal Y}^{ij}$ is reminiscent of
a membrane winding coordinate in exceptional field theory.
It transforms as
\eq{
     \delta_{\mathit\Lambda} {\cal Y}^{ij} = i[\mathit\Lambda,{\cal
      Y}^{ij}]+ [\mathit\Lambda,{\cal X}^i,{\cal X}^j]\,,\qquad\qquad
    \delta_{\mathit\Xi} {\cal Y}^{ij} =-2i \big[\mathit\Xi^{[i}, {\cal X}^{j]}\big]\,
  }
under the two kinds of gauge transformations.  
Due to the gauge covariance of $ {\cal X}^i$ one can now
add more gauge invariant terms to the action, like for instance
a mass-term
\eq{
       S_{\rm IKKT}= -&{\frac1 4} {\rm tr}\Big( \big( i[ {\cal X}^i,
         {\cal X}^i] -  {\cal Y}^{ij} \big) \big( i[ {\cal X}_i, {\cal
           X}_j] -  {\cal Y}_{ij}\big)\Big)
         + m_{ij}  {\rm tr}\Big(  [{\cal X}^i,  {\cal X}^j] \Big)   \,.       
       }
 Note that this changes the equation of motion for ${\cal X}^i$ but
 does not influence the equation of motion for ${\cal Y}^{ij}$ which
 is still the fake-flatness condition
 \eq{
             i[ {\cal X}^i, {\cal X}^j] -  {\cal Y}^{ij}=0\,.
 }      
We are not exploring these theories further in this work.

\subsection{Adding fermionic matter}

It would be interesting to know whether 
the bosonic gauge theory  of $L^{\rm cub}_2$
discussed so far allows for a supersymmetric extension.
Here we restrict ourselves to provide just a few steps towards
introducing fermionic matter in the theory.
In order to avoid too much discussion of spinors and gamma-matrices
in various dimensions, here we restrict ourselves to $D=10$.

We start at the lowest level and introduce potential gaugino superpartners to the
gauge fields $A_\mu$ and $a_\mu$.
This is  a pair of Majorana-Weyl (MW)
spinors $\Theta\in V_B$ and $\theta\in V_C$ transforming
in the adjoint representation of  $L^{\rm cub}_2$.
As usual, we combine them in a total field $\mathit{\Theta}=\Theta+\theta$,
so that their gauge variation can be compactly written as
\eq{
           \delta_{\mathit{\Lambda}} \mathit{\Theta} =i[\mathit{\Lambda},\mathit{\Theta}]\,.
}
Carrying out  a similar analysis as for the bosonic term, it turns out that
the (corrected) covariant derivative takes the form
\eq{
       \hat{\mathcal{D}}_\mu \mathit{\Theta}=\partial_\mu \mathit{\Theta} - i[{\cal} A_\mu,\mathit{\Theta}]
       +\mathit{\Psi}_\mu\,,
}
where we also had to introduce a new spin-3/2 background field $\mathit{\Psi}=\Psi+\psi\in
V_B\oplus V_C$.
Then, the covariant derivative transforms covariantly, if the background field transforms as
\eq{
  \label{psitrafo}
        \delta_{\mathit{\Lambda}}  \mathit{\Psi}_{\mu}=i[\mathit{\Lambda},\mathit{\Psi}_\mu] + [\mathit{\Lambda},{\mathcal A}_\mu,\mathit{\Theta}]\,
}
under zero-form gauge transformations and as
\eq{
       \delta_{\mathit{\Xi}}  {\mathit{\Psi}}_{\mu}=- i [\mathit{\Xi}_\mu,\mathit{\Theta}] 
}
under one-form gauge transformations.
The action
\eq{
            S=\int d^{10} x\; \Big( &-{\frac{1}{4}}{\rm tr}\big(\hat{\cal
              F}_{\mu\nu}\, \hat{\cal F}^{\mu\nu} \big)
            -{\frac{1}{2}}{\rm tr}\big(\overline{\mathit{\Theta}} \Gamma^\mu \hat{\mathcal{D}}_\mu \mathit{\Theta} \big)\Big)\,.
}
is thus gauge invariant leading to the fermionic equations of motion
\eq{
                          \Gamma^\mu \hat{\mathcal{D}}_\mu \mathit{\Theta} =0\,,\qquad
                                    \mathit\Theta=0\,,
                                  }
 where the second relation follows from the variation of
 $\mathit{\Psi}_{\mu}$. This implies also that $\mathit{\Psi}_{\mu}=0$ so
 that again the equations of motion trivialize.

The next  step is to extend the action
by  the kinetic term of the spin-3/2 field $\mathit{\Psi}_\mu$
so that one can define an action like
\eq{
            S=\int d^{10} x\; \Big(&-{\frac{1}{4}}{\rm tr}\big(\hat{\cal
              F}_{\mu\nu}\, \hat{\cal F}^{\mu\nu} \big) -{\frac{1}{6}}{\rm tr}\big(\hat{\cal
              H}_{\mu\nu\rho}\, \hat{\cal H}^{\mu\nu\rho} \big)\\
              &-{\frac{1}{2}}{\rm tr}\big(\overline{\mathit{\Theta}} \Gamma^\mu \hat{\mathcal{D}}_\mu \mathit{\Theta} \big)
            -{\frac{1}{3}}{\rm tr}\big(\overline{\mathit{\Psi}}_\mu \Gamma^{\mu\nu\rho} \hat{\mathcal{D}}_\nu \mathit{\Psi}_\rho \big)\Big)\,.
}
Recall that the spin-3/2 fermion itself transforms
as \eqref{psitrafo} and therefore not covariantly so that the kinetic term for
$\mathit{\Psi}_\mu$ is not gauge invariant. Due to the issues already encountered in the pure bosonic sector of the theory, we will not attempt to tackle the issue of a supersymmetric extension with our formalism in the present work.

\section{Conclusion}

Motivated by the still unsolved problem  of concretely introducing $M5$-branes
at a fundamental level in M-theory or Matrix theory, respectively, 
we have taken a new approach to utilize 
cubic matrices in a physical context. The main new
ingredient was that we have considered non-associative algebras
of both bi- \textit{and} cubic matrices, allowing a well defined computational
framework, in which a non-trivial ternary product arose
as the Jacobiator of the three cubic matrices.
We have pointed out that this provides a concrete example
of a two-term $L_\infty$ algebra containing a fundamental
identity involving  a three- and a two-bracket.

Moreover, we have a taken a  first few steps towards
formulating a physical theory based on such cubic matrices,
namely a generalization of Yang-Mills theory.
In a bottom-up approach we were explicitly constructing
the appearing gauge theory, which was essentially
reproducing results already reported in the literature for
generic two-term $L_\infty$ algebras.
Hence, one could view our construction  as a concrete realization
of such theories, which however have  a couple
of non-standard features.
The equations
of motion seem to be similar to Chern-Simons theory rather
than to Yang-Mills theory and, relatedly,
there is no limit in which we recover usual Yang-Mills theory.
Moreover, the addition of fermionic matter also faced
some non-trivial obstacles. On a positive note, we were able to provide some toy-examples of topological theories, where our formalism could be naturally applied.

Some of these generic issues were already
reported before, but 
this does not necessarily mean  that these are really fundamental
problems but should rather be viewed as providing the ground  for
future research. 
One might also contemplate whether
there exist other approaches to implement
these $L^{\rm cub}_2$ algebras into gauge or  gravity theories.
Maybe their natural physical application is in the description
of the six-dimensional theory on the $M5$-brane \cite{Samann:2017sxo}.
Even more speculative, there could also exist more general 
algebras of higher index matrices in which some
of the issues raised here are absent. This has the potential
to also connect to the structure of tensor hierarchies
appearing in gauged maximal supergravity and  Exceptional Field Theory.

As indicated in the introduction, our initial motivation for this project
was to find a  generalization of the BFSS Matrix Model based
on such cubic matrices. 
However, from where we now stand,
some more research is needed to arrive at such
a theory. We hope to readress this issue in future works.

\paragraph{Acknowledgments:}
The work of R.B. is supported  by the Deutsche Forschungsgemeinschaft (DFG, German Research Foundation) under Germany’s Excellence Strategy – EXC-2094 – 390783311.

\vspace{0.5cm}


\appendix

\section{Definition of L\texorpdfstring{$_\infty$}{TEXT} algebras} 
\label{app_linfty}
  
In this appendix we  briefly review the definition of an L$_\infty$ algebra.
It can be considered as a  generalized Lie algebra where one has not only a two-product, the commutator, but more general multilinear $n$-products with $n$ inputs
\eq{
\ell_n: \qquad \quad V^{\otimes n} &\rightarrow V \\
X_1, \dots , X_n &\mapsto \ell_n(X_1, \dots , X_n) \, , 
}
defined on a graded vector space $V = \bigoplus_k V_k$, where $k$ denotes the grading.
These products are graded antisymmetric 
\eq{ 
\label{permuting}
\ell_n (\dots, X_1,X_2, \dots) = (-1)^{1+ {\rm deg}(X_1) {\rm deg}( X_2)} \, \ell_n (\dots, X_2,X_1, \dots )\,,
}
with 
\eq{
      {\rm deg}\big( \, \ell_n(X_1,\ldots,X_n)\, \big)=n-2+\sum_{i=1}^n  {\rm deg}(X_i)\,.
}
 The set of higher products  $\ell_n$ define an L$_\infty$ algebra, if they satisfy the
 infinitely many relations
\eq{
\label{linftyrels}
{\cal J}_n(X_1,\ldots, X_n):=\sum_{i + j = n + 1 } &(-1)^{i(j-1)}
\sum_\sigma  (-1)^\sigma
\, \chi (\sigma;X) \; \\
 &\ell_j \big( \;
\ell_i (X_{\sigma(1)}\; , \dots , X_{\sigma(i)} )\, , X_{\sigma(i+1)} , \dots ,
X_{\sigma(n)} \, \big) = 0 \, .
}
The permutations are restricted to the ones with
\eq{ 
\label{restrictiononpermutation}
\sigma(1) < \cdots < \sigma(i)\, , \qquad \sigma(i+1) < \cdots < \sigma(n)\,,
}
and the sign $\chi(\sigma; x) = \pm 1$ can be read off from \eqref{permuting}. 
The first relations ${\cal J}_n$ with $n=1,2,3,\ldots$ can be schematically written as 
\eq{
  \label{linftyrela}
&{\cal J}_1 = \ell_1 \ell_1 \, , \qquad {\cal J}_2 = \ell_1 \ell_2 - \ell_2 \ell_1 \,,  \qquad {\cal J}_3 =
\ell_1 \ell_3 + \ell_2 \ell_2 + \ell_3 \ell_1 \,, \\[0.1cm] &
 {\cal J}_4 = \ell_1 \ell_4 - \ell_2 \ell_3 + \ell_3 \ell_2 - \ell_4 \ell_1 \, , 
}
from which one can deduce the scheme for ${\cal J}_{n>4}$. More concretely, the first L$_\infty$ relations read
\eq{   \label{ininftyrel1}
\ell_1\big( \,  \ell_1   (X) \, \big) &= 0\,, \\
\ell_1 \big( \, \ell_2(X_1, X_2)\,\big) &=  \ell_2\big(\,  \ell_1 (X_1) , X_2 \, \big) + (-1)^{X_1} \ell_2\big(\, X_1, \ell_1 (X_2) \, \big) \, ,
}
revealing that $\ell_1$ must be a nilpotent derivation with respect to
$\ell_2$, i.e. that in particular the Leibniz rule is satisfied.
Denoting $(-1)^{X_i}=(-1)^{{\rm deg}(X_i)}$ the full relation $ {\cal J}_3 $ reads
\begin{eqnarray}    \label{ininftyrel2}
       0\!\!\!&=&\!\!\! \phantom{} \ell_1\big(\ell_3(X_1,X_2,X_3)\, \big)+\ell_2\big(\ell_2(X_1,X_2),X_3\, \big)+(-1)^{(X_2+X_3)X_1}
     \ell_2\big(\ell_2(X_2,X_3),X_1\, \big)\nonumber \\[0.1cm]
   &&+(-1)^{(X_1+X_2)X_3}
     \ell_2\big(\ell_2(X_3,X_1),X_2\, \big)+\ell_3\big(\ell_1(X_1),X_2,X_3\, \big) \\[0.2cm]
     &&+(-1)^{X_1}\ell_3\big(X_1,\ell_1(X_2),X_3\, \big)
+(-1)^{X_1+X_2}\ell_3\big(X_1,X_2,\ell_1(X_3)\, \big)\, \nonumber
\end{eqnarray}   
meaning that 
the Jacobi identity for the $\ell_2$ product is mildly violated by
$\ell_1$-exact expressions.

Furthermore there exists the notion of a cyclic \textit{graded} inner product on a given $L_\infty$ algebra. This is a non-degenerate map $(\cdot, \cdot ): V \times V \to \mathbb{R}$ that is  graded symmetric, i.e.
\begin{equation}\label{gradedinnprod}
\begin{aligned}
    (X_1,X_2) &= (-1)^{X_1X_2}(X_2,X_1)\\
    (\ell_n(X_1,\dots,X_n),X_0) &= (-1)^{n+X_0(X_1+\dots+X_n)}(\ell_n(X_0,X_1,\dots,X_{n-1}),X_n)\,.
\end{aligned}
\end{equation}
Note that such an inner product need not exist for a given algebra,
cf. \cite{Ritter:2015ymv} and references therein.

\vspace{0.5cm}
\bibliography{references} 
\bibliographystyle{utphys}

\end{document}